\documentclass[prl,superscriptaddress,reprint,amsmath,amssymb]{revtex4-2}

\usepackage{bm}
\usepackage{epsfig}
\usepackage{graphicx}
\usepackage{color}
\usepackage[colorlinks]{hyperref}
\usepackage[english]{babel}
\usepackage{float}
\usepackage{empheq}
\usepackage[bbgreekl]{mathbbol}
\usepackage{units}
\usepackage[normalem]{ulem}
\usepackage{dcolumn}

\begin{document}

\author{A.~N.\ Kamenskii}
\affiliation{Experimentelle Physik 2, Technische Universität Dortmund, Dortmund, D-44221, Germany}

\author{V.~O.\ Kozlov}
\affiliation{Spin Optics Laboratory, St.\,Petersburg State University, 198504 St.\,Petersburg, Russia}
\affiliation{Photonics Department, St.\,Petersburg State University, Peterhof, 198504 St.\,Petersburg, Russia}

\author{N.~S.\ Kuznetsov}
\affiliation{Spin Optics Laboratory, St.\,Petersburg State University, 198504 St.\,Petersburg, Russia}
\affiliation{Photonics Department, St.\,Petersburg State University, Peterhof, 198504 St.\,Petersburg, Russia}

\author{I.~I.\ Ryzhov}
\affiliation{Photonics Department, St.\,Petersburg State University, Peterhof, 198504 St.\,Petersburg, Russia}
\affiliation{Spin Optics Laboratory, St.\,Petersburg State University, 198504 St.\,Petersburg, Russia}

\author{G.~G.\ Kozlov}
\affiliation{Spin Optics Laboratory, St.\,Petersburg State University, 198504 St.\,Petersburg, Russia}
\affiliation{Solid State Physics Department, St.\,Petersburg State University, Peterhof, 198504 St.\,Petersburg, Russia}

\author{M.\ Bayer}
\affiliation{Experimentelle Physik 2, Technische Universität Dortmund, Dortmund, D-44221, Germany}
\affiliation{Ioffe Institute, Politekhnycheskaya 26, St. Petersburg, 194021, Russian Federation}

\author{A.\ Greilich}
\affiliation{Experimentelle Physik 2, Technische Universität Dortmund, Dortmund, D-44221, Germany}

\author{V.~S.\ Zapasskii}
\affiliation{Spin Optics Laboratory, St.\,Petersburg State University, 198504 St.\,Petersburg, Russia}

\begin{abstract}
We show that in cubic crystals with anisotropic impurity centers the sum of squares of the magnetic resonance (EPR) frequencies  is invariant with respect to the magnetic field direction.   The connection between such an invariant and the \textit{g}-tensor components of the impurity is derived for different types of centers. The established regularity is confirmed experimentally for the spin-noise spectra of a CaF$_2$-Nd$^{3+}$ crystal. We show how this property of the EPR spectra can be efficiently used for the assignment of paramagnetic centers in cubic crystals.
\end{abstract}

\title{Invariants in the paramagnetic resonance spectra of impurity crystals}

\maketitle

\textit{Introduction} - Dielectric crystals with paramagnetic impurities are known to be classic objects of optical and electron paramagnetic resonance (EPR) spectroscopy and have remained popular materials of present-day photonics and optoelectronics~\cite{Guillot-Noel,Goldschmidt13,D2,D1}. The advantages of the impurity crystals are related to the fact that their properties can be significantly modified by small amounts of dopants without affecting the structure and macroscopic symmetry of the crystal~\cite{Sp1,Sp4,Sp5}. For many applications, dielectric crystals doped with {\it paramagnetic} impurities acquire  unique magnetic (spin-related) properties~\cite{KAPLYANSKII19871,Thiel,Handbook,Osborne}. It is important that the local symmetry of the impurity centers in many cases appears to be lower than that of the crystal host.
The anisotropy of the impurity centers, however, is not revealed in macroscopic properties of the crystal:
 cubic crystals remain magnetically and optically  isotropic, uniaxial crystals remain isotropic in the plane orthogonal to their axes~\footnote{These statements are valid for linear properties of the crystals.}, etc~\cite{PP,zap,Nye}.

The hidden symmetry of the impurity centers can be revealed using experimental methods not restricted by the effects of linear response. The most efficient among them is the method of EPR spectroscopy that makes it possible to examine the magnetic characteristics of a center in great detail. In crystals with anisotropic paramagnetic impurities with a large number of magnetically nonequivalent centers, the EPR spectra appear to be rather complex and not so simple for deciphering. Still, it is clear that the total ``smooth'' angular behavior of the linear magnetic susceptibility of the crystal (described by a second-order characteristic surface) imposes certain restrictions not only on the spatial arrangement of the centers in the crystalline matrix, but also on the angular behavior of their EPR resonances. When applied to cubic crystals with anisotropic impurity centers, which will be the main object of our research, it means, that the combination of all the EPR frequencies of the impurities have to provide invariance of the crystal magnetization with respect to the magnetic field direction.

In this Letter, we attract attention to the fact that the connection between the magnetization of an impurity crystal and its EPR frequencies is essentially different for a field of fixed direction varied in magnitude and for the field of fixed magnitude varied in direction. In the former case, both the magnetization and the EPR frequency vary linearly with the magnetic field, so that the EPR frequency can be immediately used as a measure of the magnetization. In the latter case, the EPR frequencies vary only due to a change of the effective $g$-factors of the anisotropic centers. As a result, the magnetization of the crystal, quadratically dependent on the impurity’s $g$-factor, varies as the EPR frequency squared. This fact, in combination with the magnetic isotropy of the cubic crystal, allows one to obtain an important invariant relation between EPR frequencies of the impurity centers, that is useful for interpretation of the EPR spectra in these systems. We show that this invariant is given by the sum of squares of EPR frequencies of all magnetically nonequivalent centers of the same type and is determined numerically by the values of the impurity’s \textit{g}-tensor components.

It is noteworthy that in classical EPR spectrometers the resonances are determined by the strength of magnetic fields when the spin precession frequency coincides with the fixed frequency of an applied AC field. In this respect, the method of spin noise (SN) spectroscopy (see, e.g., Refs.~\cite{Zap0,M_ller_2010,Zap1,Sinitsyn0}), that does not imply application of any AC field, allows one to obtain in a single-shot measurement the panoramic EPR spectrum of the crystal as a function of frequency, with all the frequencies needed to calculate the above invariant. We encountered the problem of deciphering multiline EPR spectra of impurity centers in cubic crystals in our studies of rare-earth-activated crystals using the method of SN spectroscopy~\cite{kamen}. These data are used here to illustrate the applicability of the proposed algorithm for assigning the EPR peaks to particular impurity centers in a cubic crystal.

\textit{Theoretical background} - To establish the origin of individual magnetic peaks, we begin with development of the theoretical background, which would allow one to relate the experimentally observed Larmor resonance frequencies to the total magnetization of the crystal in a magnetic field. Consider first the case of a cubic crystal with anisotropic paramagnetic impurity centers. The simplest way to model such a crystal goes as follows. Let us place the impurity at an arbitrary point of the elementary cell of the crystal and apply all transformations of the cubic group of symmetry. Thus we will find all points of the elementary cell equivalent to the first one from the viewpoint of symmetry. Since we have no grounds to give a preference to some of them, we assume that all these sites are filled with impurity ions uniformly. One can easily see that the number of orientationally distinguishable sites will depend on the local symmetry of the crystal in the chosen point: the number of sites of higher symmetry will be smaller than that of lower symmetry. In particular, for the cubic crystal in question, this number may be 3 (tetragonal centers), 4 (trigonal centers), 6 (rhombic centers), and 12 or 24 (centers of lower symmetry). Each of these numbers (denote it $N$) corresponds to the number of magnetically-nonequivalent centers of the appropriate symmetry. We assume the total concentration of the impurity centers to be small, so that the crystal, as a whole, keeps its cubic symmetry. In this case, in the  high-temperature limit, the tensor $\chi$ of the static magnetic susceptibility, that describes the linear relation between the magnetization of the crystal and the applied magnetic field $\bf B$ (${\bf M}=\chi{\bf B}$), is a scalar, $\chi_{ik}=\chi_0\delta_{ik}$.

Let us number the groups of magnetically equivalent centers by the superscript $\alpha=1,2,...,N$ and denote the concentration of the centers of each group by $n$. The magnetization $\bf M$ of the system is equal to the sum of the contributions ${\bf M}^\alpha$ of all groups of centers, with each of them being described by its susceptibility tensor $\chi^\alpha$:
\begin{equation}
M_i=\sum_{\alpha=1}^N M^\alpha_i=\sum_{\alpha=1}^N \chi^\alpha_{ik}B_k \Rightarrow\chi_{ik}=\sum_{\alpha=1}^N\chi_{ik}^\alpha
\end{equation}
i.e., the tensor $\chi$ of the crystal is the sum of tensors of all the groups.
For consistency of our further analysis, we present the calculation of the tensor $\chi^\alpha$~\cite{AK,AB}. The hamiltonian of an arbitrary center of the group is given by the scalar product of the operator of magnetic moment of the center $\bf m$ and the magnetic field vector $\bf B$: $H^\alpha=-m_iB_i$. The operator of magnetic moment is related to that of angular momentum (spin) $\bf S$ through the symmetric $g$-tensor:
\begin{equation}
 m_i=\mu g_{ik}^\alpha S_k
\end{equation}
Here, $\mu$ is the Bohr magneton. We will consider our paramagnetic centers as two-level systems described by the effective spin
$S = 1/2$ \footnote{This assumption is not fundamental – the calculation presented below can be easily generalized to arbitrary spin.}. In this case, the spin operator matrices $S_i$ turn into Pauli matrices, and the matrix $H^\alpha$ of the Hamiltonian of the center acquires the form:
\begin{equation}
H^\alpha=-m_iB_i=-\mu g^\alpha_{ik}S_kB_i\equiv -\mu h^{\alpha}_iS_i,\hskip 5mm
\label{41}
\end{equation}
$$
|h^\alpha|^2
 =|B|^2\bigg [ (g^\alpha_1) ^2 \cos^2\theta_1+(g^\alpha_2) ^2 \cos^2\theta_2+(g^\alpha_3) ^2 \cos^2\theta_3\bigg ],
 $$
where we introduced the effective field $h^\alpha$, and expressed its magnitude squared through the direction cosines ($\cos\theta_i,i=1,2,3$) of the magnetic field in the coordinate frame where the tensor $g^\alpha$ is diagonal with principal values $g^\alpha_i, i=1,2,3$.
Assuming the temperature $T$ so high that $\mu|h^\alpha |\ll k_BT$ ($k_B$ is the Boltzman constant), we can present the equilibrium density matrix $\rho^\alpha$ in the form:
\begin{equation}
\rho^\alpha={\frac{1-\beta H^\alpha}{2}},\hskip10mm \beta\equiv 1/k_BT.
\end{equation}
Then, for the mean value of the $i$-th projection of the magnetization, we can write the following chain of equalities:
\begin{equation}
M_i^\alpha=n\hbox{Sp }\rho^\alpha m_i ={n\beta\over 2} \hbox{Sp } H^\alpha m_i=
 {n\mu^2\beta\over 2} B_j g^\alpha_{jl} g_{ik}^\alpha \overbrace {\hbox{Sp }S_lS_k}^{\delta_{lk}/2}
\label{43}
\end{equation}
with $\hbox{Sp }$ being the trace of the matrix. Here, we used the known properties of the Pauli matrices. Now, taking into account that the tensor $g^\alpha$ is symmetric ($g^\alpha_{ik}=g^\alpha_{ki})$, we can continue the chain (\ref{43}):
\begin{equation}
M_i^\alpha=
 {n\mu^2\beta\over 4} g_{ik}^\alpha g^\alpha_{kj} B_j =
 {n\mu^2\beta\over 4} [g^\alpha]_{ij}^2 B_j =\chi^\alpha_{ij}B_j.
\end{equation}

Using this relationship, we come to the conclusion that the tensor $\chi^\alpha$ of the static magnetic susceptibility of the $\alpha$-th group and the magnetic susceptibility tensor of the whole crystal are given by the expressions:
\begin{equation}
\chi^\alpha={n\mu^2\beta\over 4}[g^\alpha ]^2,\hskip10mm
\chi= {n\mu^2\beta\over 4}\sum_{\alpha=1}^N [g^\alpha ]^2.
\label{46}
\end{equation}
Note that, as was already mentioned, for the magnetically isotropic crystal, $\chi_{ik}=\chi_0\delta_{ik}$, where $\chi_0$ is a scalar. Therefore, for such crystals, the sum of squares of the $g$-tensors of the group of paramagnetic centers is a scalar, which we denote by $g^2_0I$ ($I$ is the unity matrix):
\begin{equation}
\sum_{\alpha=1}^N[g^\alpha]^2_{ik}=g^2_{0}\delta_{ik}
\label{47}
\end{equation}

For the crystal with paramagnetic impurities occupying positions that cross over into each other under the symmetry transformations of the crystal, the principal values of the $g$-tensors of all centers are the same ($g_i^\alpha\equiv g_i, i=1,2,3$), and the $g$-tensors of each group differ only by orientations of their axes. In this case, the factor $g_0^2$ entering Eq. (\ref{47}) can be expressed through principal values of the $g$-tensors ($g_1, g_2,$ and $ g_3$) by taking the trace of the right- and left-hand sides of this equation:
\begin{equation}
g_{0}^2={N\over 3}[g_1^2+g_2^2+g_3^2]
\label{48}
\end{equation}

Now, let us turn to calculation of the EPR frequencies of the above crystal. The $\alpha$-th group of centers provides a peak in the EPR spectrum at the frequency $\omega_\alpha$ determined by the difference of eigenvalues of Hamiltonian (\ref{41}): $\hbar\omega_\alpha= \mu |h^\alpha|$.
Using definition (\ref{41}) for the components $h^\alpha _i$ of the effective field, we can write the following expression for the sum of squares of the EPR frequencies:
\begin{equation}
\sum_{\alpha=1}^N\omega_\alpha ^2={\mu^2\over\hbar^2}\sum_{\alpha=1}^Nh_i^\alpha h_i^\alpha=
{\mu^2\over\hbar^2}\sum_{\alpha=1}^Ng_{ik}^\alpha g_{ij}^\alpha B_k B_j =
\label{10a}
\end{equation}
$$
={\mu^2\over\hbar^2}\sum_{\alpha=1}^N[g^\alpha]^2_{kj} B_k B_j.
$$
Taking into account Eqs.~(\ref{47}) and (\ref{48}), we eventually have:
\begin{equation}
\sum_{\alpha=1}^N\omega_\alpha ^2= {N\over 3} \bigg ({\mu B\over\hbar}\bigg )^2 [g_1^2+g_2^2+g_3^2].
\label{11}
\end{equation}

Thus, we come to the conclusion that the sum of squares of the EPR frequencies created by all magnetically-nonequivalent centers of the same type in a cubic crystal is invariant with respect to the magnetic field direction and is determined numerically by the sum of squares of the $g$-tensor components of these centers. For a crystal with impurity centers of different types, the total number of EPR lines increases, but the EPR spectra of centers of each type obey their own invariant relationships. In the general case, the above invariants do not allow one to unambiguously identify EPR spectra of a cubic crystal, but may essentially simplify this problem.

Note now that Eq.~(\ref{10a}) can be written in the form:
 \begin{equation}
 \sum_{\alpha=1}^N\omega_\alpha^2={4\over n\beta \hbar^2}({\bf M,B}),\hskip5mm {\bf M}\equiv \sum_{\alpha=1}^N{\bf M}^\alpha,
 \label{12}
 \end{equation}
which shows that, in the general case, regardless of the crystal symmetry, the sum of squares of the EPR frequencies of paramagnetic centers is determined by the {\it projection} of the total magnetization $\bf M$ of these centers onto the magnetic field $\bf B$. This fact establishes a direct relation between the EPR frequencies of the impurity crystal of arbitrary symmetry and its magnetization. In conformity with the crystal symmetry, the above sum becomes invariant with respect to the magnetic field direction not only for cubic crystals, but also for uniaxial crystals provided that the magnetic field is rotated in its ``equatorial'' plane.

\begin{figure}[h!t]
\includegraphics[width=\linewidth]{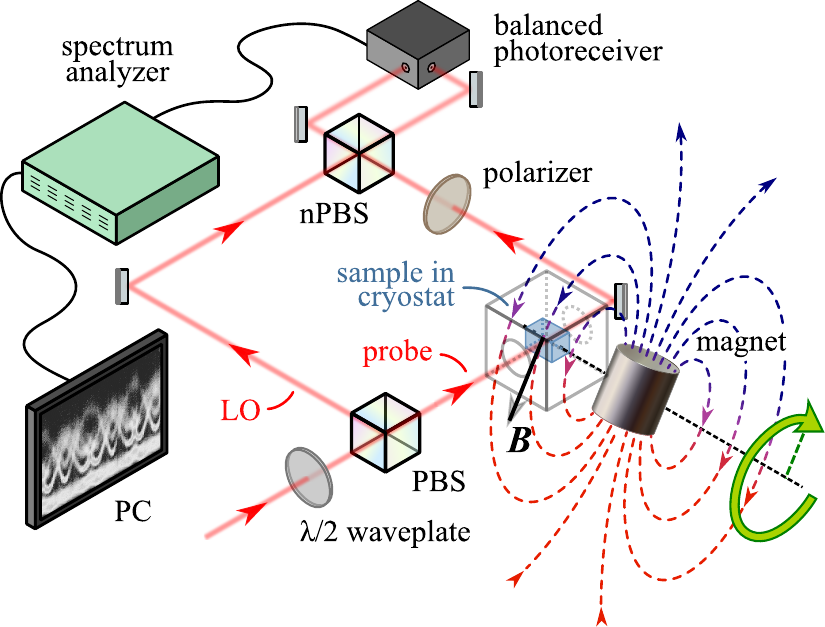}
\caption{Schematic of the homodyne-detection setup with a rotating magnetic field of constant strength. The laser beam is split at the input polarizing beamsplitter (PBS) of the Mach-Zehnder interferometer into  two parts: the signal (probe) and local oscillator (LO). The transmitted and scattered photons of the probe  have orthogonal linear polarizations. The polarizer is used to filter out the transmitted part. The passed scattered light and the light of the LO are matched to interfere on the 50:50 non-polarizing beamsplitter (nPBS). The balanced photoreceiver measures the interference signal. The AC part of the signal is analyzed using a spectrum analyzer.}
\label{fig1}
\end{figure}

\begin{figure}[h!t]
 \includegraphics[width=\linewidth]{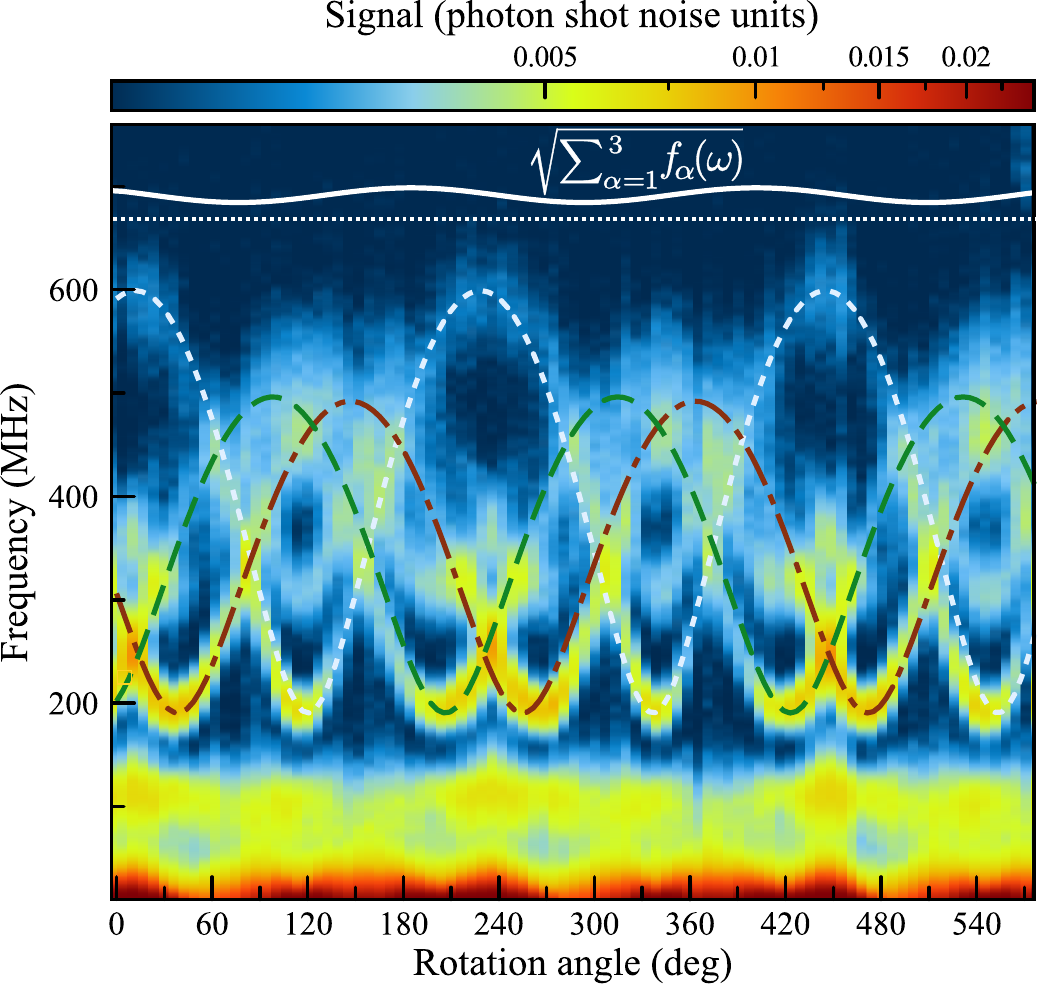}
 \caption{Experimental angular dependence of the SN spectrum of the crystal CaF$_2$-Nd$^{3+}$ (0.1 mol. $\%$), obtained by rotation of the applied magnetic field (created by an outer permanent magnet). The dashed curves represent approximations of the angular variations of the SN resonance frequencies calculated by the relationship $\hbar \omega/\mu B_m=\sqrt{a +b\cos [2\alpha]+c \sin [2\alpha]}$.
 The wavy (solid) and straight (dotted) lines show, respectively,
 experimental and calculated by Eq.~(\ref{11}) angular dependencies of the sum of squares of the SN resonance frequencies.
 }
 \label{fig2}
 \end{figure}

\textit{Experimental illustration} - To confirm the above conclusions, we took advantage of the SN spectroscopy method that allows one to measure simultaneously all the resonance frequencies (spin precession frequencies) observed in a fixed magnetic field.

The optical arrangement of the setup, schematically presented in Fig.~\ref{fig1}, was similar to that described in Refs.~\cite{Petrov,kamen}. As an object of study, we chose the sample CaF$_2$-Nd$^{3+}$ (0.1 mol.\%) used in our recent work~\cite{kamen} to demonstrate the applicability of the SN spectroscopy to dielectric crystals with paramagnetic impurities. The laser wavelength was fixed at 862.69\,nm, which corresponds to one of the absorption lines of the Nd$^{3+}$ centers in this crystal~\cite{kamen}. A plane-parallel plate of the crystal with arbitrarily oriented crystallographic axes was held in a cryostat at a temperature of $\sim$ 6\,K. Unlike conventional measurements of the SN spectra, performed in a fixed magnetic field applied orthogonal to the optical axis, the present experiments implied measuring the SN spectra at varying orientation of the magnetic field with respect to the crystal. We solved this task, as we believe, in the simplest way, using a strong permanent magnet. The disk-like magnet magnetized along its cylindrical axis was placed outside the cryostat, as shown in Fig.~\ref{fig1}, and could be rotated around the axis $L$ lying in the plane of the disk and passing through the center of the sample. As a result, the magnetic field created by the magnet on the sample, could be rotated in the plane normal to the axis $L$, remaining the same in magnitude. This arrangement allowed us to make measurements of the SN spectra at different orientations of the fixed magnetic field and, in addition, to monitor the continuous transformation of the SN spectra with  rotation of  the magnetic field.

Figure~\ref{fig2} shows the results of the measurements in a colormap format. There are several features that allowed us to decipher the picture fairly easily. First of all, the Nd$^{3+}$ ion in the CaF$_2$ crystal, replacing the divalent cation Ca$^{2+}$, typically occupies a tetragonal position with the F$^-$ ion compensating the excess impurity charge. The components of the ground-state $g$-tensor of this center are known to be g$_{\parallel}$ = 4.412 and g$_{\perp}$ = 1.301~\cite{AK}. Regarding the general pattern of the angular dependence of the SN resonance frequencies, one notices that, for the magnetic field rotating in a fixed plane, the picture should be 180$^\circ$-periodic and reach its minima (corresponding to the component g$_{\perp}$) three times per each 180$^\circ$. The frequencies corresponding to all these minima should be evidently the same. These three curves, related to three groups of magnetically nonequivalent centers can be well distinguished in the experimental picture.

To find out the values of the invariants, we have performed an accurate quantitative analysis of the experimental data. The mentioned minima of the angular dependencies, the same for all three groups of the centers and approximately equal to $\omega_{min}=2\pi\cdot 185\cdot 10^6$ rad/sec (see Fig.~\ref{fig2}), allowed us to evaluate more accurately the magnetic field $B_m$ created by the permanent magnet on the sample. It is given by the relationship $B_m=\hbar\omega_{min}/g_\bot \mu=10$\,mT (here $g_\bot=1.301$). As can be shown, the behavior of the presented angular dependencies is described by sinusoidal dependence of the resonance frequencies {\it squared} in the form $[\hbar \omega/\mu B_m]^2=a +b\cos [2\alpha]+c \sin [2\alpha]$, where $\alpha$ is the angle specifying the orientation of the magnetic field in the plane of its rotation. Results of the approximation of the experimental angular dependencies by this equation are presented in Fig.~\ref{fig2} by the dashed curves. 
In the figure, we also show the angular behavior of the sum $\sum_{i=1}^3[\hbar \omega_i/\mu B_m]^2$ (which is expected to be constant) and its calculated value $2g_\bot^2+g_\|^2$ (see Eq.~(\ref{11})). As seen from Fig.~\ref{fig2}, the measured sum $\sum_{i=1}^3[\hbar \omega_i/\mu B_m]^2$ is, within the experimental accuracy, invariant with respect to the magnetic field direction and corresponds well to its calculated value shown by the white dotted horizontal line.
	
These results show unambiguously that the three chosen SN-resonance peaks belong to different magnetically nonequivalent groups of the same tetragonal Nd$^{3+}$ centers.

At the same time, the experimental SN resonance spectra presented in Fig.~\ref{fig2} reveal some additional peaks with a different angular dependence. We did not intend to examine accurately these dependencies, but we can conclude that these peaks undoubtedly belong to Nd$^{3+}$ centers (because of the strong selectivity of the method in the optical channel) with other values of the \textit{g}-tensor components~\cite{kamen}.

\textit{Conclusion} - To conclude, we have shown that in crystals with anisotropic paramagnetic centers the sum of  squares of the EPR frequencies is proportional to the projection of the magnetization of the centers onto the magnetic field. In cubic crystals, in view of their isotropy, the magnetisation is aligned along the applied field, and this sum remains invariant upon rotation. The value of this invariant is easily expressed through the impurity’s \textit{g}-tensor components and can be used for identification of the centers. In uniaxial crystals, this quantity should be invariant to rotation of the field in the plane orthogonal to the crystal axis. In crystals of arbitrary symmetry, this sum follows the general pattern of the crystal magnetization and indicates its orientation with respect to the external magnetic field. The angular behavior of this sum for all magnetically nonequivalent centers of any (but the same) type reflects the general magnetic anisotropy (or isotropy) of the crystal through EPR frequencies of its impurities.

\begin{acknowledgments}
We highly appreciate the financial support from the Deutsche Forschungsgemeinschaft in the frame of the International Collaborative Research Center TRR 160 (Project A5) and the Russian Foundation for Basic Research (Grant No. 19-52-12054). The authors from Russian side acknowledge the Saint Petersburg State University for the research Grant No. 73031758.
\end{acknowledgments}


\begin{thebibliography}{24}%
\makeatletter
\providecommand \@ifxundefined [1]{%
 \@ifx{#1\undefined}
}%
\providecommand \@ifnum [1]{%
 \ifnum #1\expandafter \@firstoftwo
 \else \expandafter \@secondoftwo
 \fi
}%
\providecommand \@ifx [1]{%
 \ifx #1\expandafter \@firstoftwo
 \else \expandafter \@secondoftwo
 \fi
}%
\providecommand \natexlab [1]{#1}%
\providecommand \enquote  [1]{``#1''}%
\providecommand \bibnamefont  [1]{#1}%
\providecommand \bibfnamefont [1]{#1}%
\providecommand \citenamefont [1]{#1}%
\providecommand \href@noop [0]{\@secondoftwo}%
\providecommand \href [0]{\begingroup \@sanitize@url \@href}%
\providecommand \@href[1]{\@@startlink{#1}\@@href}%
\providecommand \@@href[1]{\endgroup#1\@@endlink}%
\providecommand \@sanitize@url [0]{\catcode `\\12\catcode `\$12\catcode
  `\&12\catcode `\#12\catcode `\^12\catcode `\_12\catcode `\%12\relax}%
\providecommand \@@startlink[1]{}%
\providecommand \@@endlink[0]{}%
\providecommand \url  [0]{\begingroup\@sanitize@url \@url }%
\providecommand \@url [1]{\endgroup\@href {#1}{\urlprefix }}%
\providecommand \urlprefix  [0]{URL }%
\providecommand \Eprint [0]{\href }%
\providecommand \doibase [0]{https://doi.org/}%
\providecommand \selectlanguage [0]{\@gobble}%
\providecommand \bibinfo  [0]{\@secondoftwo}%
\providecommand \bibfield  [0]{\@secondoftwo}%
\providecommand \translation [1]{[#1]}%
\providecommand \BibitemOpen [0]{}%
\providecommand \bibitemStop [0]{}%
\providecommand \bibitemNoStop [0]{.\EOS\space}%
\providecommand \EOS [0]{\spacefactor3000\relax}%
\providecommand \BibitemShut  [1]{\csname bibitem#1\endcsname}%
\let\auto@bib@innerbib\@empty
\bibitem [{\citenamefont {Guillot-No\"{e}l}\ \emph {et~al.}(2009)\citenamefont
  {Guillot-No\"{e}l}, \citenamefont {Goldner}, \citenamefont {Beaudoux},
  \citenamefont {Lejay}, \citenamefont {Amari}, \citenamefont {Walther},
  \citenamefont {Rippe}, \citenamefont {Kr\"{o}ll}, \citenamefont
  {Chaneli\`{e}re},\ and\ \citenamefont {Gou\"{e}t}}]{Guillot-Noel}%
  \BibitemOpen
  \bibfield  {author} {\bibinfo {author} {\bibfnamefont {O.}~\bibnamefont
  {Guillot-No\"{e}l}}, \bibinfo {author} {\bibfnamefont {P.}~\bibnamefont
  {Goldner}}, \bibinfo {author} {\bibfnamefont {F.}~\bibnamefont {Beaudoux}},
  \bibinfo {author} {\bibfnamefont {J.}~\bibnamefont {Lejay}}, \bibinfo
  {author} {\bibfnamefont {A.}~\bibnamefont {Amari}}, \bibinfo {author}
  {\bibfnamefont {A.}~\bibnamefont {Walther}}, \bibinfo {author} {\bibfnamefont
  {L.}~\bibnamefont {Rippe}}, \bibinfo {author} {\bibfnamefont
  {S.}~\bibnamefont {Kr\"{o}ll}}, \bibinfo {author} {\bibfnamefont
  {T.}~\bibnamefont {Chaneli\`{e}re}},\ and\ \bibinfo {author} {\bibfnamefont
  {J.~L.~L.}\ \bibnamefont {Gou\"{e}t}},\ }in\ \href@noop {} {\emph {\bibinfo
  {booktitle} {CLEO/Europe and EQEC 2009 Conference Digest}}}\ (\bibinfo
  {publisher} {Optical Society of America},\ \bibinfo {year}
  {2009})\BibitemShut {NoStop}%
\bibitem [{\citenamefont {Goldschmidt}\ \emph {et~al.}(2013)\citenamefont
  {Goldschmidt}, \citenamefont {Peters}, \citenamefont {Polyakov},
  \citenamefont {Migdall}, \citenamefont {Beavan},\ and\ \citenamefont
  {Sellars}}]{Goldschmidt13}%
  \BibitemOpen
  \bibfield  {author} {\bibinfo {author} {\bibfnamefont {E.~A.}\ \bibnamefont
  {Goldschmidt}}, \bibinfo {author} {\bibfnamefont {J.}~\bibnamefont {Peters}},
  \bibinfo {author} {\bibfnamefont {S.~V.}\ \bibnamefont {Polyakov}}, \bibinfo
  {author} {\bibfnamefont {A.~L.}\ \bibnamefont {Migdall}}, \bibinfo {author}
  {\bibfnamefont {S.~E.}\ \bibnamefont {Beavan}},\ and\ \bibinfo {author}
  {\bibfnamefont {M.~J.}\ \bibnamefont {Sellars}},\ }in\ \href@noop {} {\emph
  {\bibinfo {booktitle} {Frontiers in Optics 2013}}}\ (\bibinfo  {publisher}
  {Optical Society of America},\ \bibinfo {year} {2013})\ p.\ \bibinfo {pages}
  {LTu1G.2}\BibitemShut {NoStop}%
\bibitem [{\citenamefont {{Zhong}}\ and\ \citenamefont {{Goldner}}(2019)}]{D2}%
  \BibitemOpen
  \bibfield  {author} {\bibinfo {author} {\bibfnamefont {T.}~\bibnamefont
  {{Zhong}}}\ and\ \bibinfo {author} {\bibfnamefont {P.}~\bibnamefont
  {{Goldner}}},\ }\href@noop {} {\bibfield  {journal} {\bibinfo  {journal}
  {Nanophotonics}\ }\textbf {\bibinfo {volume} {8}},\ \bibinfo {eid} {185}
  (\bibinfo {year} {2019})}\BibitemShut {NoStop}%
\bibitem [{\citenamefont {Paniagua-Dominguez}\ \emph
  {et~al.}(2020)\citenamefont {Paniagua-Dominguez}, \citenamefont
  {Luk'yanchuk},\ and\ \citenamefont {Kuznetsov}}]{D1}%
  \BibitemOpen
  \bibfield  {author} {\bibinfo {author} {\bibfnamefont {R.}~\bibnamefont
  {Paniagua-Dominguez}}, \bibinfo {author} {\bibfnamefont {B.}~\bibnamefont
  {Luk'yanchuk}},\ and\ \bibinfo {author} {\bibfnamefont {A.~I.}\ \bibnamefont
  {Kuznetsov}},\ }in\ \href@noop {} {\emph {\bibinfo {booktitle} {Dielectric
  Metamaterials}}},\ \bibinfo {series and number} {Woodhead Publishing Series
  in Electronic and Optical Materials},\ \bibinfo {editor} {edited by\ \bibinfo
  {editor} {\bibfnamefont {I.}~\bibnamefont {Brener}}, \bibinfo {editor}
  {\bibfnamefont {S.}~\bibnamefont {Liu}}, \bibinfo {editor} {\bibfnamefont
  {I.}~\bibnamefont {Staude}}, \bibinfo {editor} {\bibfnamefont
  {J.}~\bibnamefont {Valentine}},\ and\ \bibinfo {editor} {\bibfnamefont
  {C.}~\bibnamefont {Holloway}}}\ (\bibinfo  {publisher} {Woodhead
  Publishing},\ \bibinfo {year} {2020})\ pp.\ \bibinfo {pages}
  {73--108}\BibitemShut {NoStop}%
\bibitem [{\citenamefont {Marfunin}(1979)}]{Sp1}%
  \BibitemOpen
  \bibfield  {author} {\bibinfo {author} {\bibfnamefont {A.~S.}\ \bibnamefont
  {Marfunin}},\ }\bibinfo {title} {Electron paramagnetic resonance},\ in\
  \href@noop {} {\emph {\bibinfo {booktitle} {Spectroscopy, Luminescence and
  Radiation Centers in Minerals}}}\ (\bibinfo  {publisher} {Springer Berlin
  Heidelberg},\ \bibinfo {address} {Berlin, Heidelberg},\ \bibinfo {year}
  {1979})\ pp.\ \bibinfo {pages} {76--118}\BibitemShut {NoStop}%
\bibitem [{\citenamefont {Liu}(2005)}]{Sp4}%
  \BibitemOpen
  \bibfield  {author} {\bibinfo {author} {\bibfnamefont {G.}~\bibnamefont
  {Liu}},\ }\bibinfo {title} {Electronic energy level structure},\ in\
  \href@noop {} {\emph {\bibinfo {booktitle} {Spectroscopic Properties of Rare
  Earths in Optical Materials}}},\ \bibinfo {editor} {edited by\ \bibinfo
  {editor} {\bibfnamefont {R.}~\bibnamefont {Hull}}, \bibinfo {editor}
  {\bibfnamefont {J.}~\bibnamefont {Parisi}}, \bibinfo {editor} {\bibfnamefont
  {R.~M.}\ \bibnamefont {Osgood}}, \bibinfo {editor} {\bibfnamefont
  {H.}~\bibnamefont {Warlimont}}, \bibinfo {editor} {\bibfnamefont
  {G.}~\bibnamefont {Liu}},\ and\ \bibinfo {editor} {\bibfnamefont
  {B.}~\bibnamefont {Jacquier}}}\ (\bibinfo  {publisher} {Springer Berlin
  Heidelberg},\ \bibinfo {address} {Berlin, Heidelberg},\ \bibinfo {year}
  {2005})\ pp.\ \bibinfo {pages} {1--94}\BibitemShut {NoStop}%
\bibitem [{\citenamefont {Jacquier}\ \emph {et~al.}(2005)\citenamefont
  {Jacquier}, \citenamefont {Bigot}, \citenamefont {Guy},\ and\ \citenamefont
  {Jurdyc}}]{Sp5}%
  \BibitemOpen
  \bibfield  {author} {\bibinfo {author} {\bibfnamefont {B.}~\bibnamefont
  {Jacquier}}, \bibinfo {author} {\bibfnamefont {L.}~\bibnamefont {Bigot}},
  \bibinfo {author} {\bibfnamefont {S.}~\bibnamefont {Guy}},\ and\ \bibinfo
  {author} {\bibfnamefont {A.~M.}\ \bibnamefont {Jurdyc}},\ }\bibinfo {title}
  {Rare earth doped confined structures for lasers and amplifiers},\ in\
  \href@noop {} {\emph {\bibinfo {booktitle} {Spectroscopic Properties of Rare
  Earths in Optical Materials}}},\ \bibinfo {editor} {edited by\ \bibinfo
  {editor} {\bibfnamefont {R.}~\bibnamefont {Hull}}, \bibinfo {editor}
  {\bibfnamefont {J.}~\bibnamefont {Parisi}}, \bibinfo {editor} {\bibfnamefont
  {R.~M.}\ \bibnamefont {Osgood}}, \bibinfo {editor} {\bibfnamefont
  {H.}~\bibnamefont {Warlimont}}, \bibinfo {editor} {\bibfnamefont
  {G.}~\bibnamefont {Liu}},\ and\ \bibinfo {editor} {\bibfnamefont
  {B.}~\bibnamefont {Jacquier}}}\ (\bibinfo  {publisher} {Springer Berlin
  Heidelberg},\ \bibinfo {address} {Berlin, Heidelberg},\ \bibinfo {year}
  {2005})\ pp.\ \bibinfo {pages} {430--461}\BibitemShut {NoStop}%
\bibitem [{\citenamefont {Kaplyanskii}\ and\ \citenamefont
  {Ryskin}(1987)}]{KAPLYANSKII19871}%
  \BibitemOpen
  \bibfield  {author} {\bibinfo {author} {\bibfnamefont {A.}~\bibnamefont
  {Kaplyanskii}}\ and\ \bibinfo {author} {\bibfnamefont {A.}~\bibnamefont
  {Ryskin}},\ }in\ \href@noop {} {\emph {\bibinfo {booktitle} {Spectroscopy of
  Solids Containing Rare Earth Ions}}},\ \bibinfo {series} {Modern Problems in
  Condensed Matter Sciences}, Vol.~\bibinfo {volume} {21},\ \bibinfo {editor}
  {edited by\ \bibinfo {editor} {\bibfnamefont {A.}~\bibnamefont
  {Kaplyanskii}}\ and\ \bibinfo {editor} {\bibfnamefont {R.}~\bibnamefont
  {Macfarlane}}}\ (\bibinfo  {publisher} {Elsevier},\ \bibinfo {year} {1987})\
  pp.\ \bibinfo {pages} {1--12}\BibitemShut {NoStop}%
\bibitem [{\citenamefont {Thiel}\ \emph {et~al.}(2011)\citenamefont {Thiel},
  \citenamefont {B{\"o}ttger},\ and\ \citenamefont {Cone}}]{Thiel}%
  \BibitemOpen
  \bibfield  {author} {\bibinfo {author} {\bibfnamefont {C.}~\bibnamefont
  {Thiel}}, \bibinfo {author} {\bibfnamefont {T.}~\bibnamefont {B{\"o}ttger}},\
  and\ \bibinfo {author} {\bibfnamefont {R.}~\bibnamefont {Cone}},\ }\href@noop
  {} {\bibfield  {journal} {\bibinfo  {journal} {Journal of Luminescence}\
  }\textbf {\bibinfo {volume} {131}},\ \bibinfo {pages} {353} (\bibinfo {year}
  {2011})}\BibitemShut {NoStop}%
\bibitem [{\citenamefont {Goldner}\ \emph {et~al.}(2015)\citenamefont
  {Goldner}, \citenamefont {Ferrier},\ and\ \citenamefont
  {Guillot-No{\"e}l}}]{Handbook}%
  \BibitemOpen
  \bibfield  {author} {\bibinfo {author} {\bibfnamefont {P.}~\bibnamefont
  {Goldner}}, \bibinfo {author} {\bibfnamefont {A.}~\bibnamefont {Ferrier}},\
  and\ \bibinfo {author} {\bibfnamefont {O.}~\bibnamefont {Guillot-No{\"e}l}},\
  }\href@noop {} {\bibfield  {journal} {\bibinfo  {journal} {Handbook on The
  Physics and Chemistry of Rare Earths}\ }\textbf {\bibinfo {volume} {46}},\
  \bibinfo {pages} {1} (\bibinfo {year} {2015})}\BibitemShut {NoStop}%
\bibitem [{\citenamefont {Osborne}(2017)}]{Osborne}%
  \BibitemOpen
  \bibfield  {author} {\bibinfo {author} {\bibfnamefont {I.~S.}\ \bibnamefont
  {Osborne}},\ }\href@noop {} {\bibfield  {journal} {\bibinfo  {journal}
  {Science}\ }\textbf {\bibinfo {volume} {357}},\ \bibinfo {pages} {1366}
  (\bibinfo {year} {2017})}\BibitemShut {NoStop}%
\bibitem [{Note1()}]{Note1}%
  \BibitemOpen
  \bibinfo {note} {These statements are valid for linear properties of the
  crystals.}\BibitemShut {Stop}%
\bibitem [{\citenamefont {Feofilov}\ and\ \citenamefont
  {Kaplyanskii}(1962)}]{PP}%
  \BibitemOpen
  \bibfield  {author} {\bibinfo {author} {\bibfnamefont {P.~P.}\ \bibnamefont
  {Feofilov}}\ and\ \bibinfo {author} {\bibfnamefont {A.~A.}\ \bibnamefont
  {Kaplyanskii}},\ }\href@noop {} {\bibfield  {journal} {\bibinfo  {journal}
  {Soviet Physics Uspekhi}\ }\textbf {\bibinfo {volume} {5}},\ \bibinfo {pages}
  {79} (\bibinfo {year} {1962})}\BibitemShut {NoStop}%
\bibitem [{\citenamefont {Zapasskii}(1973)}]{zap}%
  \BibitemOpen
  \bibfield  {author} {\bibinfo {author} {\bibfnamefont {V.~S.}\ \bibnamefont
  {Zapasskii}},\ }\href@noop {} {\bibfield  {journal} {\bibinfo  {journal}
  {Soviet Physics Uspekhi}\ }\textbf {\bibinfo {volume} {15}},\ \bibinfo
  {pages} {2367} (\bibinfo {year} {1973})}\BibitemShut {NoStop}%
\bibitem [{\citenamefont {Nye}(1985)}]{Nye}%
  \BibitemOpen
  \bibfield  {author} {\bibinfo {author} {\bibfnamefont {J.}~\bibnamefont
  {Nye}},\ }\href@noop {} {\emph {\bibinfo {title} {Physical properties of
  crystals : their representation by tensors and matrices}}}\ (\bibinfo
  {publisher} {Oxford University Press},\ \bibinfo {year} {1985})\BibitemShut
  {NoStop}%
\bibitem [{\citenamefont {Alexandrov}\ and\ \citenamefont
  {Zapasskii}(1981)}]{Zap0}%
  \BibitemOpen
  \bibfield  {author} {\bibinfo {author} {\bibfnamefont {E.}~\bibnamefont
  {Alexandrov}}\ and\ \bibinfo {author} {\bibfnamefont {V.}~\bibnamefont
  {Zapasskii}},\ }\href@noop {} {\bibfield  {journal} {\bibinfo  {journal}
  {Sov. Phys. JETP}\ }\textbf {\bibinfo {volume} {54}} (\bibinfo {year}
  {1981})}\BibitemShut {NoStop}%
\bibitem [{\citenamefont {Müller}\ \emph {et~al.}(2010)\citenamefont
  {Müller}, \citenamefont {Oestreich}, \citenamefont {Römer},\ and\
  \citenamefont {Hübner}}]{M_ller_2010}%
  \BibitemOpen
  \bibfield  {author} {\bibinfo {author} {\bibfnamefont {G.~M.}\ \bibnamefont
  {Müller}}, \bibinfo {author} {\bibfnamefont {M.}~\bibnamefont {Oestreich}},
  \bibinfo {author} {\bibfnamefont {M.}~\bibnamefont {Römer}},\ and\ \bibinfo
  {author} {\bibfnamefont {J.}~\bibnamefont {Hübner}},\ }\href@noop {}
  {\bibfield  {journal} {\bibinfo  {journal} {Physica E: Low-dimensional
  Systems and Nanostructures}\ }\textbf {\bibinfo {volume} {43}},\ \bibinfo
  {pages} {569–587} (\bibinfo {year} {2010})}\BibitemShut {NoStop}%
\bibitem [{\citenamefont {Zapasskii}(2013)}]{Zap1}%
  \BibitemOpen
  \bibfield  {author} {\bibinfo {author} {\bibfnamefont {V.}~\bibnamefont
  {Zapasskii}},\ }\href@noop {} {\bibfield  {journal} {\bibinfo  {journal}
  {Adv. Opt. Photon.}\ }\textbf {\bibinfo {volume} {5}},\ \bibinfo {pages}
  {131} (\bibinfo {year} {2013})}\BibitemShut {NoStop}%
\bibitem [{\citenamefont {Sinitsyn}\ and\ \citenamefont
  {Pershin}(2016)}]{Sinitsyn0}%
  \BibitemOpen
  \bibfield  {author} {\bibinfo {author} {\bibfnamefont {N.}~\bibnamefont
  {Sinitsyn}}\ and\ \bibinfo {author} {\bibfnamefont {Y.}~\bibnamefont
  {Pershin}},\ }\href@noop {} {\bibfield  {journal} {\bibinfo  {journal}
  {Reports on Progress in Physics}\ }\textbf {\bibinfo {volume} {79}} (\bibinfo
  {year} {2016})}\BibitemShut {NoStop}%
\bibitem [{\citenamefont {Kamenskii}\ \emph {et~al.}(2020)\citenamefont
  {Kamenskii}, \citenamefont {Greilich}, \citenamefont {Ryzhov}, \citenamefont
  {Kozlov}, \citenamefont {Bayer},\ and\ \citenamefont {Zapasskii}}]{kamen}%
  \BibitemOpen
  \bibfield  {author} {\bibinfo {author} {\bibfnamefont {A.~N.}\ \bibnamefont
  {Kamenskii}}, \bibinfo {author} {\bibfnamefont {A.}~\bibnamefont {Greilich}},
  \bibinfo {author} {\bibfnamefont {I.~I.}\ \bibnamefont {Ryzhov}}, \bibinfo
  {author} {\bibfnamefont {G.~G.}\ \bibnamefont {Kozlov}}, \bibinfo {author}
  {\bibfnamefont {M.}~\bibnamefont {Bayer}},\ and\ \bibinfo {author}
  {\bibfnamefont {V.~S.}\ \bibnamefont {Zapasskii}},\ }\href@noop {} {\bibfield
   {journal} {\bibinfo  {journal} {Phys. Rev. Research}\ }\textbf {\bibinfo
  {volume} {2}},\ \bibinfo {pages} {023317} (\bibinfo {year}
  {2020})}\BibitemShut {NoStop}%
\bibitem [{AK(1964)}]{AK}%
  \BibitemOpen
  in\ \href@noop {} {\emph {\bibinfo {booktitle} {Electron Paramagnetic
  Resonance}}},\ \bibinfo {editor} {edited by\ \bibinfo {editor} {\bibfnamefont
  {S.}~\bibnamefont {Al'tshuler}}\ and\ \bibinfo {editor} {\bibfnamefont
  {B.}~\bibnamefont {Kozyrev}}}\ (\bibinfo  {publisher} {Academic Press},\
  \bibinfo {year} {1964})\ p.\ \bibinfo {pages} {369}\BibitemShut {NoStop}%
\bibitem [{\citenamefont {Abragam}\ and\ \citenamefont {Bleaney}(1970)}]{AB}%
  \BibitemOpen
  \bibfield  {author} {\bibinfo {author} {\bibfnamefont {A.}~\bibnamefont
  {Abragam}}\ and\ \bibinfo {author} {\bibfnamefont {B.}~\bibnamefont
  {Bleaney}},\ }\href@noop {} {\emph {\bibinfo {title} {Electron paramagnetic
  resonance of transition metal ions}}}\ (\bibinfo  {publisher} {Clerendon
  Press},\ \bibinfo {year} {1970})\BibitemShut {NoStop}%
\bibitem [{Note2()}]{Note2}%
  \BibitemOpen
  \bibinfo {note} {This assumption is not fundamental – the calculation
  presented below can be easily generalized to arbitrary spin.}\BibitemShut
  {Stop}%
\bibitem [{\citenamefont {Petrov}\ \emph {et~al.}(2018)\citenamefont {Petrov},
  \citenamefont {Kamenskii}, \citenamefont {Zapasskii}, \citenamefont {Bayer},\
  and\ \citenamefont {Greilich}}]{Petrov}%
  \BibitemOpen
  \bibfield  {author} {\bibinfo {author} {\bibfnamefont {M.~Y.}\ \bibnamefont
  {Petrov}}, \bibinfo {author} {\bibfnamefont {A.~N.}\ \bibnamefont
  {Kamenskii}}, \bibinfo {author} {\bibfnamefont {V.~S.}\ \bibnamefont
  {Zapasskii}}, \bibinfo {author} {\bibfnamefont {M.}~\bibnamefont {Bayer}},\
  and\ \bibinfo {author} {\bibfnamefont {A.}~\bibnamefont {Greilich}},\
  }\href@noop {} {\bibfield  {journal} {\bibinfo  {journal} {Phys. Rev. B}\
  }\textbf {\bibinfo {volume} {97}},\ \bibinfo {pages} {125202} (\bibinfo
  {year} {2018})}\BibitemShut {NoStop}%
\end{thebibliography}
\bibliographystyle{apsrev}

\end{document}